\documentclass[times,twocolumn,final]{elsarticle}
\makeatletter

\usepackage{framed,multirow}

\usepackage{amssymb}
\usepackage{latexsym}
\usepackage{pifont}
\usepackage{url}
\usepackage{xcolor}

\usepackage{hyperref}
\usepackage{halloweenmath}
\usepackage{cleveref}
\usepackage{marvosym}
\usepackage{dsfont}

\usepackage{tikz}
\newcommand*\circled[1]{\tikz[baseline=(char.base)]{
            \node[shape=circle,draw,inner sep=0.75pt] (char) {#1};}}

\let\oldpercent\%
\renewcommand{\%}{\scalebox{0.85}{\oldpercent}}

\definecolor{newcolor}{rgb}{.8,.349,.1}

\begin{document}

\begin{frontmatter}

\title{Registration-Enhanced Segmentation Method for Prostate Cancer in Ultrasound Images}%




\cortext[*]{Corresponding author: \\ 
Mirabela Rusu; Email: mirabela.rusu@stanford.edu \\ Geoffrey A. Sonn; Email: gsonn@stanford.edu}
  
\author[a,b]{Shengtian Sang}
\author[a,c]{Hassan Jahanandish}
\author[d]{Cynthia Xinran Li}
\author[c,e]{Indrani Bhattachary}
\author[a,c]{Jeong Hoon Lee}
\author[a]{Lichun Zhang}
\author[a,c]{Sulaiman Vesal}
\author[c]{Pejman Ghanouni}
\author[c]{Richard Fan}
\author[a,c]{Geoffrey A. Sonn}
\author[a,c]{Mirabela Rusu}

\address[a]{Stanford University, Department of Radiology, 300 Pasteur Drive, Stanford, 94305, California, USA}
\address[b]{University of Miami, Department of Computer Science, Coral Gables, 33124, Florida, USA}
\address[c]{Stanford University, Department of Urology, 300 Pasteur Drive, Stanford, 94305, California, USA}
\address[d]{Stanford University, Institute of Computational and Mathematical Engineering, 475 Via Ortega, Stanford, 94305, California, USA}
\address[e]{Dartmouth College, Department of Biomedical Data Science, 1 Medical Center Drive, Lebanon, 03756, New Hampshire, USA}

\begin{abstract}
Prostate cancer is a major cause of cancer-related deaths in men, where early detection greatly improves survival rates. Although MRI-TRUS fusion biopsy offers superior accuracy by combining MRI's detailed visualization with TRUS's real-time guidance, it is a complex and time-intensive procedure that relies heavily on manual annotations, leading to potential errors. To address these challenges, we propose a fully automatic MRI-TRUS fusion-based segmentation method that identifies prostate tumors directly in TRUS images without requiring manual annotations. Unlike traditional multimodal fusion approaches that rely on naive data concatenation, our method integrates a registration-segmentation framework to align and leverage spatial information between MRI and TRUS modalities. This alignment enhances segmentation accuracy and reduces reliance on manual effort. Our approach was validated on a dataset of 1,747 patients from Stanford Hospital, achieving an average Dice coefficient of 0.212, outperforming TRUS-only (0.117) and naive MRI-TRUS fusion (0.132) methods, with significant improvements (p $<$ 0.01). This framework demonstrates the potential for reducing the complexity of prostate cancer diagnosis and provides a flexible architecture applicable to other multimodal medical imaging tasks.
\end{abstract}


\end{frontmatter}


\section{Introduction}
\begin{figure*}
    \centering
    \includegraphics[width=0.7\linewidth]{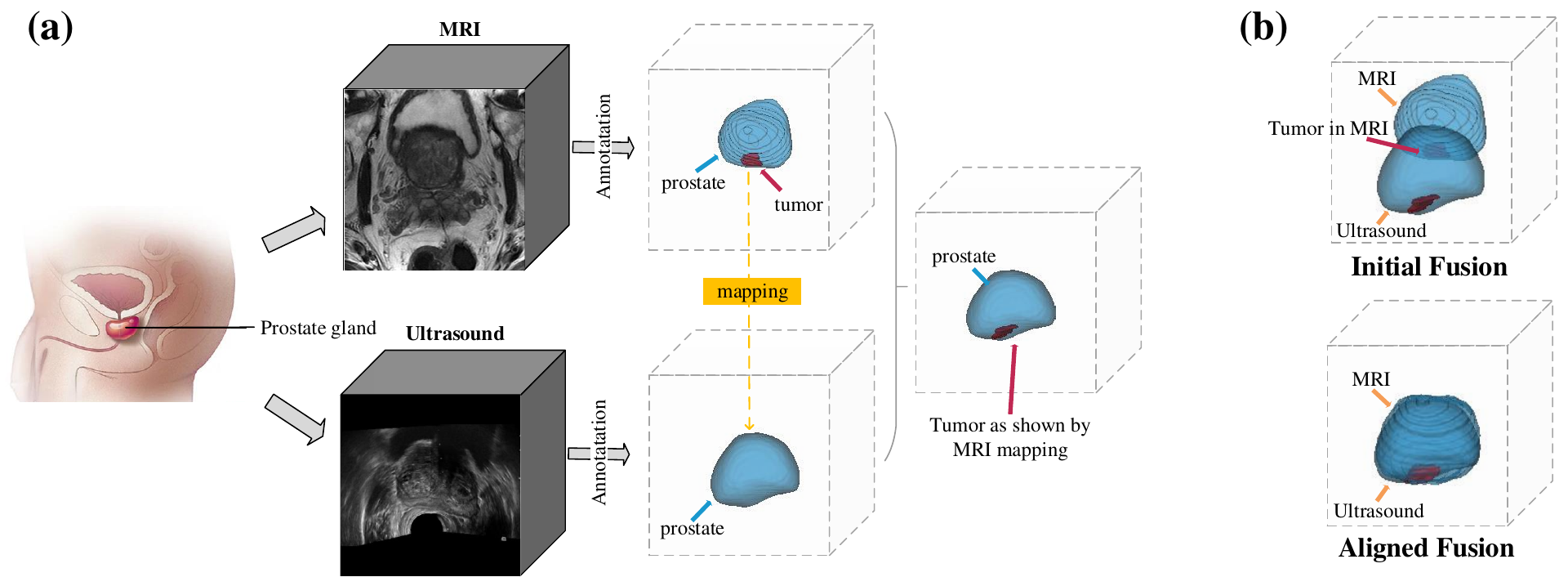}
    \caption{(a) Illustration of the MRI-TRUS fusion-based biopsy process. The patient first undergoes an MRI scan, during which a physician annotates the suspicious tumor region and the prostate gland on the MRI image. During the biopsy procedure, the tumor region identified on the MRI is mapped onto the ultrasound image to guide the needle placement for sampling. (b) Illustration of the fused MRI and ultrasound images. The top part of the figure shows the MRI and ultrasound images placed in the same spatial coordinate system, highlighting the initial misalignment of prostate information between the two modalities. The bottom part displays the MRI-ultrasound fusion after alignment using our method, showing improved alignment of prostate information across both imaging modalities.}
\label{Figure introduction}
\end{figure*}

Prostate cancer is the second most common cancer and the fifth leading cause of cancer-related death among men \cite{gandaglia2021epidemiology, center2012international}. Early detection of prostate cancer is crucial for effective treatment, as patients diagnosed in an early stage can have a 5-year survival rate that exceeds 99\% \cite{brawley2012trends}. Transrectal ultrasound (TRUS)-guided biopsy is the most common method for diagnosing prostate cancer because TRUS provides real-time imaging of the prostate \cite{pokorny2014prospective}. However, due to the low signal-to-noise ratio of ultrasound images, up to 52\% of clinically significant cancer lesions may be missed during TRUS-only biopsy procedures \cite{russo2021diagnostic}. In contrast, while magnetic resonance imaging (MRI) has challenges with real-time imaging, it produces clearer images of the prostate gland and is more effective at identifying cancerous areas \cite{kobus2012prostate}. As a result, MRI-TRUS fusion biopsy is considered a more advanced technique for the diagnosis of prostate cancer \cite{das2020mri}. In clinical practice, a multiparametric MRI (mpMRI) scan is typically performed first to evaluate the likelihood of prostate cancer and to locate any suspicious lesions. Radiologists delineate the prostate and mark suspicious regions on the MRI images. These MRI images are then fused with real-time ultrasound images during the procedure. Using ultrasound as real-time guidance, urologists perform a \textbf{fusion-guided} biopsy to accurately direct the needle into the suspicious regions identified on the \cite{sonn2014target,oberlin2016diagnostic,xu2008real}, as illustrated in Fig.\ref{Figure introduction}a. As the reliability of MRI-TRUS fusion-guided biopsy continues to be demonstrated, this technique has gained widespread popularity worldwide \cite{russo2021diagnostic}. However, it remains more complex and time-consuming than the traditional \textbf{TRUS-guided} biopsy \cite{mager2017novice,lenfant2024learning}. The process often involves identifying potential cancerous areas in the MRI and annotating the prostate in both MRI and TRUS images. While advances in automation have reduced the manual effort required, significant time and expertise are still needed to ensure accurate annotation and image fusion \cite{cool2015evaluation}. Furthermore, if the physician identifies areas on the MRI that are not cancerous (false positives) or misses tumor areas (false negatives) \cite{quon2015false}, these errors may propagate to the fused TRUS images, potentially affecting biopsy accuracy. While real-time TRUS guidance allows for some degree of clinical correction based on ultrasound characteristics, the accuracy of the procedure remains heavily dependent on the quality of MRI interpretation and image fusion.\\
This work presents an automatic \textbf{MRI-TRUS fusion-based} segmentation method that can identify suspected tumor regions in TRUS images without requiring manual annotations, thereby facilitating subsequent biopsy procedures. Traditional multimodal fusion techniques often train models by simply concatenating data from different modalities at various stages \cite{karani2022review,stahlschmidt2022multimodal,gadzicki2020early,boulahia2021early}. However, such straightforward concatenation of prostate MRI and ultrasound data fails to effectively enhance tumor segmentation performance. This limitation arises because prostate-related information is frequently misaligned between the initial MRI and ultrasound images due to differences in patient positioning, imaging protocols, or inherent modality disparities (as illustrated in the upper part of Fig.\ref{Figure introduction}b). Accurate segmentation relies heavily on spatial information \cite{ghafoorian2017location}, which is not adequately preserved or utilized in naive concatenation approaches. When MRI and ultrasound images are combined for training, misaligned tumor location information from MRI often interferes with the model's predictions on TRUS data, leading to degraded performance. To address this, we propose a novel registration-segmentation multimodal fusion technique that progressively aligns prostate information from MRI and TRUS images (as shown in the bottom part of Fig.\ref{Figure introduction}b). We validated our method using data from 1,747 patients at Stanford Hospital. For the tumor segmentation task, our method achieved an average Dice coefficient of 0.212, compared to 0.117 for models trained with TRUS alone and 0.117 for models trained with naive MRI/TRUS fusion. This corresponds to relative improvements of 81.2\% and 60.6\%, respectively. These results were statistically significant (p < 0.01), demonstrating the superior performance of our approach in accurately segmenting tumor regions. 
Our main contributions are threefold: 
\begin{itemize}
\item We propose a fully automatic MRI-TRUS fusion-based segmentation method to enhance tumor identification in ultrasound images by improving segmentation accuracy. This approach eliminates the need for physicians to manually annotate tumors on MRI, significantly reducing both time and effort.
\item Our experiments reveal that simply increasing the amount of data does not always improve prostate tumor segmentation. Instead, effective utilization of raw multimodal data is crucial for enhancing AI model performance. To address this, we propose a registration-segmentation fusion approach that optimally leverages spatial and modality-specific features.
\item The proposed registration-segmentation framework is highly flexible and can be easily adapted to various registration and segmentation methods. Furthermore, it is applicable to other disease segmentation tasks involving multimodal data, such as liver cancer or brain tumor segmentation. 

\end{itemize}

\section{Related Work}
\subsection{AI in Prostate Cancer Diagnosis and Biopsy}
Ultrasound-guided biopsy is the most widely used approach due to its real-time imaging capability and relatively low cost. Grayscale transrectal ultrasound (TRUS)\cite{bhattacharya2022review} is the most commonly adopted imaging method. However, its inherently low signal-to-noise ratio makes tumor identification challenging, with more than 50\% of tumors likely to be missed\cite{choi2019comparison,azizi2018deep}. Recent advances in ultrasound-based imaging (such as shear-wave elastography\cite{woo2017shear,sigrist2017ultrasound}, color Doppler ultrasound\cite{liau2019prostate}, contrast-enhanced ultrasound\cite{ashrafi2018contrast}, and micro-ultrasound\cite{harland2021micro}) have shown promise in improving tumor clarity by providing additional functional and structural information. Despite these advancements, relatively few studies have leveraged AI technologies to detect prostate tumors by analyzing ultrasound images alone\cite{bhattacharya2022review}. Some researchers have developed various machine learning and deep learning methods to identify prostate tumors from ultrasound images\cite{wildeboer2020automated,sedghi2019deep,han2008computer,jahanandish2024deep}. However, due to the low-quality and unclear nature of prostate ultrasound images, AI-based analysis typically achieves limited performance, with Dice scores often reported to be low\cite{pensa2024evaluation}.

MRI is increasingly utilized for detecting prostate cancer\cite{bhattacharya2022review, pensa2024evaluation}. It plays a critical role in guiding MRI-ultrasound fusion biopsies and supporting treatment planning. It is widely recognized as the most sensitive noninvasive imaging modality, capable of accurately visualizing, detecting, and localizing prostate cancer. Recent advancements have shown promising results in leveraging AI for prostate cancer detection on MRI\cite{khan2021recent,ye2022medical}. These AI-driven approaches primarily aim to identify tumors on MRI scans and subsequently map the detected tumors onto ultrasound images during clinical procedures, facilitating precise guidance for biopsies.

Here, we propose an approach similar to TRUS-MRI fusion, where MRI and TRUS are used to train the model. However, instead of identifying tumors on MRI and mapping them onto TRUS, our method directly identifies tumors on TRUS images. This approach eliminates the performance degradation caused by registration errors, offering a more streamlined and accurate solution for tumor detection.

\subsection{Multimodal Fusion Method}
Multimodal fusion methods, which integrate various types of data, have become a crucial approach to improving model performance in numerous tasks\cite{stahlschmidt2022multimodal}. The primary strategies for multimodal data fusion can be categorized into early fusion, intermediate fusion, and late fusion. These methods differ mainly in the stages at which the data are combined during the model's processing. Early fusion involves concatenating different data types before entering them into the model\cite{rajalingam2017multimodality,ronen2019evaluation,albaradei2021metacancer}. This approach allows the model to learn from the combined features of the input data simultaneously\cite{venugopalan2021multimodal,he2021imageomics,he2021imageomics,mobadersany2018predicting}. Intermediate fusion, on the other hand, concatenates the features extracted by the model from the different modalities. This method leverages the model's ability to independently extract relevant features from each modality before combining them. Lastly, late fusion combines the outputs of models that process each modality separately and then feeds the concatenated outputs into another model, which makes the final decision related to the task\cite{deng2020multimodal,huang2020multimodal,huang2020multimodal}. Regardless of whether early, intermediate, or late fusion is used, the primary data fusion technique involves directly concatenating raw data or extracted features. These concatenation operations do not account for the spatial positional information of the data. In this work, we propose a feature fusion method based on data alignment, which first aligns the spatial information of multimodal data to enhance the performance of segmentation models. This approach is crucial because the spatial positioning of the multimodal data significantly impacts the segmentation performance of the model. 

\subsection{Joint Registration and Segmentation Methods}
Recently, some studies have integrated registration and segmentation to improve semantic segmentation tasks \cite{elmahdy2021joint, xu2019deepatlas, pohl2006bayesian, pham2024regsegnet}. One common strategy alternates the optimization of segmentation and registration networks. For example, DeepAtlas \cite{xu2019deepatlas} and RegSegNet \cite{pham2024regsegnet} alternate one registration step with one segmentation step in an iterative process. In these approaches, separate phases are employed, where registration and segmentation are independently trained for each epoch, with prior phases frozen and subsequent ones excluded.
Another approach treats registration and segmentation as two tasks trained simultaneously. For instance, the Cross-Stitch Network treats registration and segmentation as a multi-task learning problem, enabling parameter sharing between the tasks to enhance performance \cite{elmahdy2021joint}. Most existing methods rely heavily on deformable registration techniques, significantly increasing model complexity. For example, DeepAtlas fixes one task while training the other due to GPU memory limitations, preventing the simultaneous optimization of both tasks \cite{xu2019deepatlas}.
In this work, we adopt affine registration to reduce the complexity of the registration process. Our method uses registration primarily to align data and provide it as input for the segmentation task. This design prioritizes segmentation accuracy, leveraging registration as a supporting process to improve the quality of the input data. Experimental results demonstrate that this framework achieves improved performance and efficiency in semantic segmentation.

\section{Materials and Methods}
\subsection{Dataset Acquisition and Preprocessing}
Our study received approval from the Institutional Review Board (IRB) at Stanford University and included 1,747 patients who underwent MRI-TRUS fusion-targeted biopsy using the Artemis system. The Hitachi Hi-Vision 5500 7.5 MHz end-firing ultrasound probe was employed to acquire 3D TRUS scans by rotating the probe 200 degrees around its axis and subsequently interpolating the scans to isotropic resolution (voxel spacing: $\sim$0.5 mm). A 3 Tesla GE scanner with external 32-channel body array coils was used during the MRI examination. Axial T2-weighted MRI images (acquired using a 2D Spin Echo protocol) from each patient were resampled to the same spatial resolution ($\sim$0.5 mm) in the axial plane, with a distance of 3 mm between slices and were used in this study.

\textbf{Preprocessing}. The TRUS and MRI scans had varying voxel spacings and matrix sizes. To standardize the input data, we resampled the ultrasound and MRI images using trilinear interpolation to achieve an isotropic voxel size of 0.5 mm. This was followed by cropping and resizing the images to a fixed size of 256×256×256. Furthermore, we normalized the TRUS image intensities using z-score normalization, computed based on the mean and standard deviation of intensities within the prostate.

\textbf{Ground Truth Labels}. We derived the ground truth labels through a multi-step process. First, radiologists outlined lesions on MRI to obtain the initial cancer labels, which were subsequently mapped onto TRUS images using the non-rigid registration provided by the Artemis device. These preliminary labels were then refined using MRI-ultrasound biopsy results to correct the manually annotated tumors. Finally, an expert manually annotated the labels on the TRUS images. The dataset was randomly divided into training, validation, and test sets, consisting of 1,116, 280, and 351 patients.

\begin{figure*}
    \centering
    \includegraphics[width=0.85\linewidth]{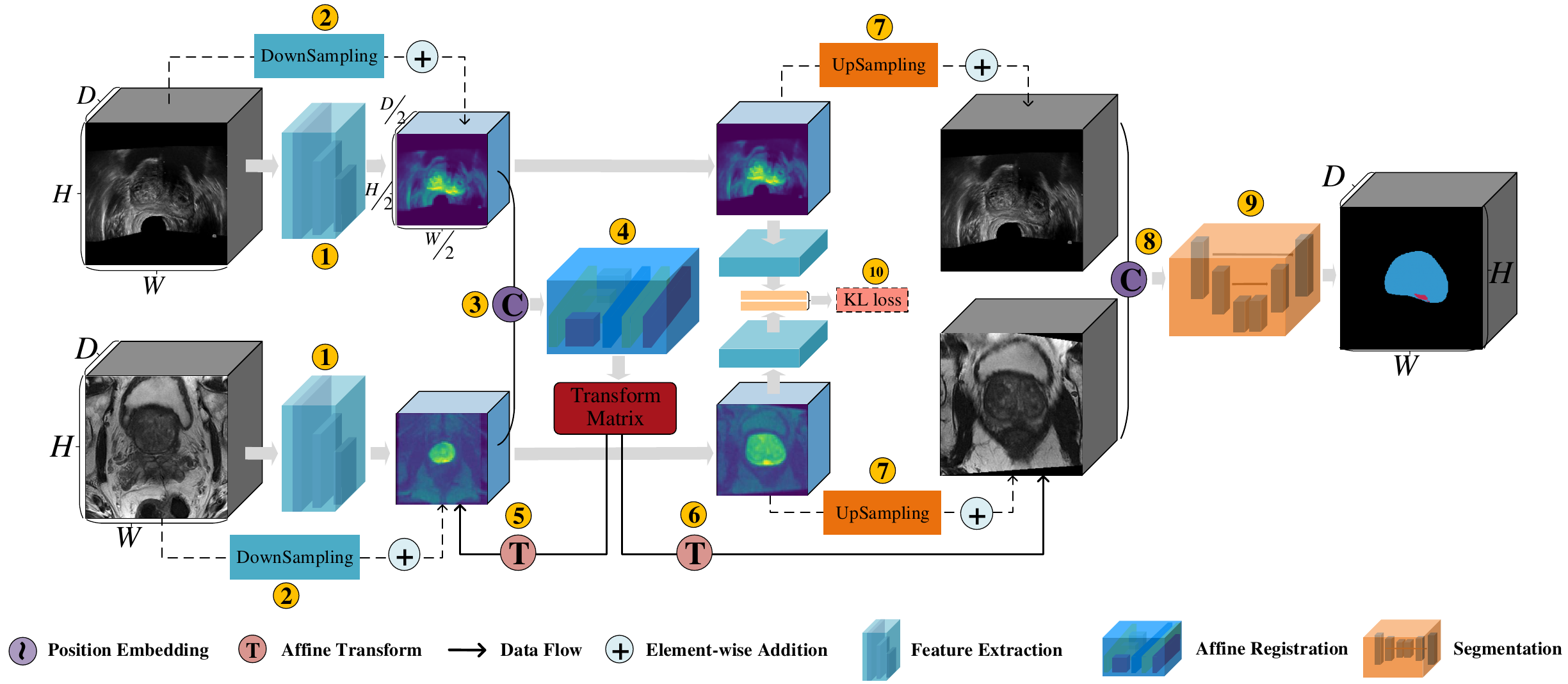}
    \caption{This figure illustrates the overall architecture of the proposed method. The numbered components represent key steps in the framework, with steps 4 and 9 corresponding to the registration and segmentation processes. The architecture incorporates positional embeddings (Step 1) and feature extraction (Steps 2 and 3) to process input images. Affine registration (Step 4) is performed to align the data, generating a transform matrix that guides the segmentation process (Step 9). Arrows indicate data flows, while element-wise addition for feature integration is denoted by the ``+'' symbol. Upsampling (Step 7) and downsampling (Step 2) further refine the outputs.}
    \label{Figure method}
\end{figure*}

\subsection{Registration-Enhanced Segmentation Method}
\label{method:regsegnet}
The main idea of this method is to integrate the registration and segmentation processes simultaneously during multimodal data training. The interference introduced by multimodal data fusion into the segmentation model is mitigated by progressively aligning the MRI and ultrasound data through the registration method. Figure \ref{Figure method} illustrates the framework of the approach, with blocks \circled{4} and \circled{9} representing the registration and segmentation methods, respectively. The input to the method comprises preprocessed MRI and ultrasound images with identical dimensions, denoted as $\boldsymbol{M}_{init}$ and $\boldsymbol{T}_{init}$, where $\boldsymbol{M}_{init} \in \mathbb{R}^{W \times H \times D}$ and $\boldsymbol{T}_{init} \in \mathbb{R}^{W \times H \times D}$.

Step \circled{1}. The feature extraction modules, consisting of two convolutional neural networks (CNNs) and activation functions, are used to extract features from both MRI and TRUS images. The stride of the first CNN is set to 2, reducing the feature size to $\mathbb{R}^{\frac{W}{2} \times \frac{H}{2} \times \frac{D}{2}}$, with the extracted features denoted as $\boldsymbol{M_{fea}}$ and $\boldsymbol{T_{fea}}$ for MRI and TRUS, respectively. This step is designed to capture prostate-related information from both modalities. In particular, we formulate this procedure as \cref{eq:step1}:

\begin{equation}
\begin{aligned}
  \boldsymbol{M}_{\text{fea}} &= Conv_1^2(Conv_1^1(\boldsymbol{M}_{init})), \\
  \boldsymbol{T}_{\text{fea}} &= Conv_2^2(Conv_2^1(\boldsymbol{T}_{init})),
\end{aligned}
\label{eq:step1}
\end{equation}
 where $\boldsymbol{M}_{fea}$ and  $\boldsymbol{T}_{fea} \in \mathbb{R}^{\frac{W}{2} \times \frac{H}{2} \times \frac{D}{2}}$.

Step \circled{2}.
Average pooling is used to downsample the original MRI and ultrasound images, reducing their dimensions to align with the feature size extracted in Step 1. The downsampled data, which preserves the original input features, is then combined with the features obtained in the first step. By combining the features extracted in steps 1 and 2, we ensure that all original features are maintained while the key features are highlighted. This process is described by the following \cref{eq:step2}:

\begin{equation}
\begin{aligned}
  \boldsymbol{M}_{\text{merge}} &= AvgPool(\boldsymbol{M}_{init})+\boldsymbol{M}_{fea}, \\
  \boldsymbol{T}_{\text{merge}} &= AvgPool(\boldsymbol{T}_{init})+\boldsymbol{T}_{fea} ,
\end{aligned}
\label{eq:step2}
\end{equation}
 where $\boldsymbol{M}_{merge}$ and  $\boldsymbol{T}_{merge} \in \mathbb{R}^{\frac{W}{2} \times \frac{H}{2} \times \frac{D}{2}}$.

Step \circled{3}.
Step 3: $\boldsymbol{M}_{merge}$ and $\boldsymbol{T}_{merge}$ are concatenated along the channel dimension and then fed into the registration module. Our registration module aims to align not only the prostate regions in MRI and TRUS images but also the features extracted from both modalities, enabling the model to learn shared critical features for better alignment and improved multimodal fusion.

\begin{figure*}
    \centering
    \includegraphics[width=0.85\linewidth]{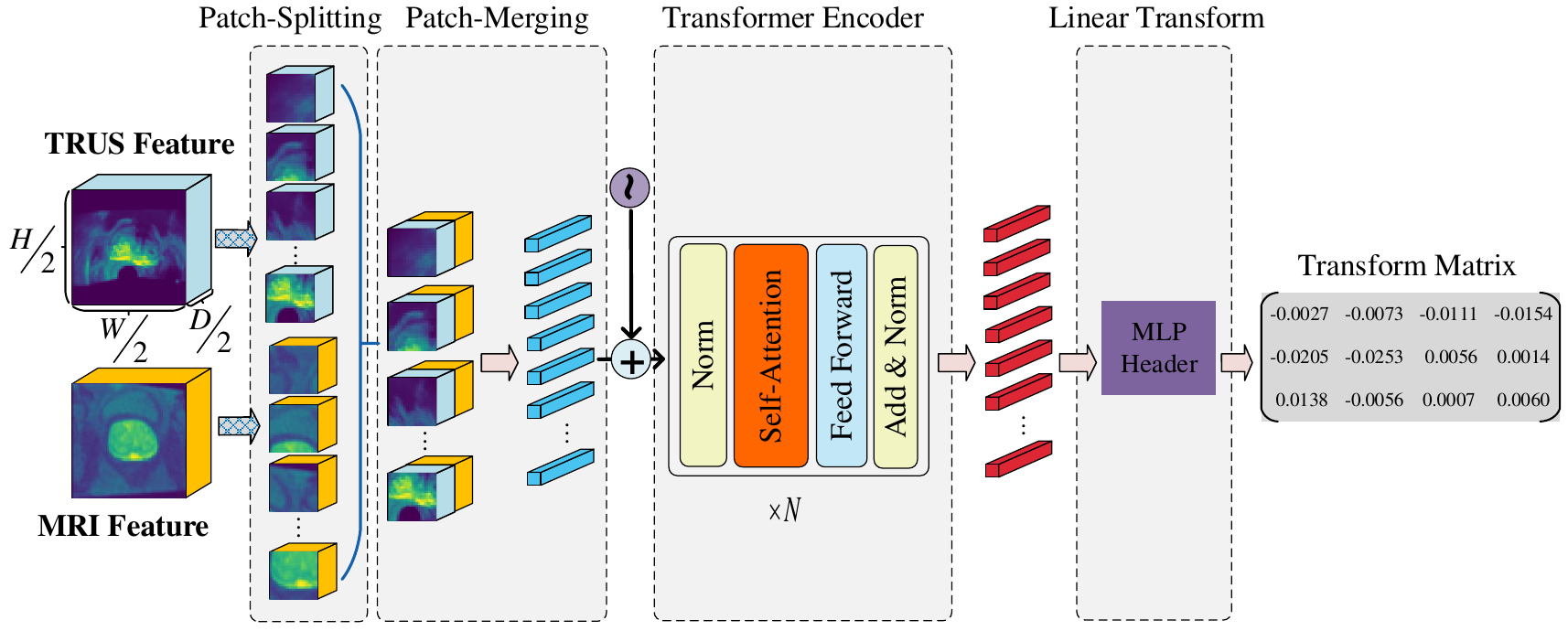}
    \caption{The architecture of the registration method adopted in our approach. The input TRUS and MRI features are initially divided into equal-sized patches through a patch-splitting operation, and a patch-merging step then combines and reshapes the features. The transformed feature vectors pass through N consecutive Transformer Encoder layers, which consist of self-attention, feed-forward, and normalization blocks. Finally, the Multi-Layer Perceptron (MLP) processes the output vectors to generate the Transform Matrix, which enables affine transformations.}
    \label{Figure method registration}
\end{figure*}

Step \circled{4}.The input for the registration module consists of $\boldsymbol{T}_{merge}$ and $\boldsymbol{M}_{merge}$ and the output is the affine transformation matrix $A$ which is used to align $\boldsymbol{M}_{merge}$ with $\boldsymbol{T}_{merge}$. The architecture involves patch-splitting, patch-merging and Transformer-based encoder layers\cite{vaswani2017attention} as shown in Fig.\ref{Figure method registration}. Same as ViT \cite{dosovitskiy2020image}, the inputs $\boldsymbol{M}_{merge}$ and $\boldsymbol{T}_{merge}$ are first splitted into non-overlapping image patches by using a sliding window with stride $k$. The input images are then reshaped into a matrix of size $\mathbb{R}^{N \times k^3}$, where $N$ represents the number of patches, calculated as ${\frac{W}{2k} \times \frac{H}{2k} \times \frac{D}{2k}}$. The patch embeddings of $\boldsymbol{M}_{merge}$ and $\boldsymbol{T}_{merge}$ are then concatenated to form a combined embedding matrix of size $\mathbb{R}^{N \times 2k^3}$. A linear layer is subsequently applied to transform this concatenated embedding into $\textbf{Z}^0$ of size $\mathbb{R}^{N \times C}$. Transformer blocks are then applied to $\textbf{Z}^0$, consisting of a multi-head self-attention (MSA) module, followed by a 2-layer MLP\cite{taud2018multilayer} with GELU non-linearity\cite{hendrycks2016gaussian}. Finally, a multi-linear layer is used on the output of encoder to generate the affine transformation matrix $\boldsymbol{A}$. This process is described by the following \cref{eq:registration}:

\begin{equation}
\begin{aligned}
\boldsymbol{Z}^{0}=\mathds{M}(\mathds{S}(\boldsymbol{T}_{\text{fea}},\boldsymbol{M}_{\text{fea}})), \\
\boldsymbol{Z}^\ell=MSA(LN(\boldsymbol{Z}^{\ell-1}))+\boldsymbol{Z}^{\ell-1}, \ell=1,2...n, 
\\
\boldsymbol{A} =MLP(LN(\boldsymbol{Z}^n)), 
\end{aligned}
\label{eq:registration}
\end{equation}

where $\boldsymbol{Z}^0$ and $\boldsymbol{Z}^l \in \mathbb{R}^{N \times C}$, $\boldsymbol{A} \in \mathbb{R}^{3 \times 4}$. $\mathds{S}$ and $\mathds{M}$ represent the patch splitting and patch merging, respectively.

Step \circled{5}. The transformation matrix $\boldsymbol{A}$ is used to transform $\boldsymbol{M}_{merge}$ to $\hat{\boldsymbol{M}}_{merge}$ which align with $\boldsymbol{T}_{merge}$. This process is described by the following \cref{eq:step 5}:

\begin{equation}
\begin{aligned}
\hat{\boldsymbol{M}}_{merge}=\mathds{T}(\boldsymbol{M}_{merge},\boldsymbol{A}), \\
\end{aligned}
\label{eq:step 5}
\end{equation}
$\mathds{T}$ represents the affine transform option, $\hat{\boldsymbol{M}}_{merge} \in \mathbb{R}^{\frac{W}{2k} \times \frac{H}{2k} \times \frac{D}{2k}}$

Step \circled{6}. This matrix is used not only to align the features extracted from $\boldsymbol{M}{init}$ and $\boldsymbol{T}_{init}$ but also to align the initial images.
\begin{equation}
\begin{aligned}
\hat{\boldsymbol{M}}_{init}=\mathds{T}(\boldsymbol{M}_{init},\boldsymbol{A}), \\
\end{aligned}
\label{eq:step 6}
\end{equation}
where $\hat{\boldsymbol{M}}_{init} \in \mathbb{R}^{W \times H \times D}$.

Step \circled{7}. $\boldsymbol{T}_{merge}$ and $\hat{\boldsymbol{M}}_{merge}$ are first upsampled to the same size of $\boldsymbol{M}_{init}$ and $\boldsymbol{T}_{init}$. Then the upsampled data are integrated into the $\boldsymbol{T}_{init}$ and $\hat{\boldsymbol{M}}_{init}$ for segmentation. This process is described by the following:
\begin{equation}
\begin{aligned}
\hat{\boldsymbol{M}}_{up}=\mathds{U}(\hat{\boldsymbol{M}}_{merge}), \\
\boldsymbol{T}_{up}=\mathds{U}(\boldsymbol{T}_{merge}), 
\end{aligned}
\label{eq:step 6}
\end{equation}
where $\mathds{U}$ indicates the upsampling operation, $\hat{\boldsymbol{M}}_{up}$ and $\boldsymbol{T}_{up}  \in \mathbb{R}^{W \times H \times D}$.

Step \circled{8}. $\boldsymbol{T}_{up}$ and $\hat{\boldsymbol{M}}_{up}$ are added to corresponding initial image modalidy $\boldsymbol{T}_{init}$ and $\hat{\boldsymbol{M}}_{init}$, respectively. Then they are concatenated along with the first dimension.

\begin{equation}
\begin{aligned}
\widehat{\hat{\boldsymbol{M}}\boldsymbol{T}}=\mathds{C}(\hat{\boldsymbol{M}}_{up}+\hat{\boldsymbol{M}}_{init},
\boldsymbol{T}_{up}+\boldsymbol{T}_{init}), 
\end{aligned}
\label{eq:step 8}
\end{equation}
$\mathds{C}$ represents the concatinate operation.

Step \circled{9}. The concatenated MRI and TRUS are input into the segmentation module, and the output is the segmentation mask of the prostate cancer $\boldsymbol{Y}$, where $\boldsymbol{Y} \in \mathbb{R}^{W \times H \times D}$.

\begin{equation}
\begin{aligned}
\boldsymbol{Y}=\boldsymbol{Seg}(\widehat{\hat{\boldsymbol{M}}\boldsymbol{T}}), 
\end{aligned}
\label{eq:step 9}
\end{equation}
$\boldsymbol{Seg}$ represents the segmentation method. In this work, we adopted U-Net as the segmentation module, as it is one of the most widely used segmentation methods and serves as a suitable choice to validate our proposed registration-segmentation framework. In practice, it can be easily replaced with other segmentation methods.

\subsection{Training Loss}
The key idea of our method is to align information from multimodal prostate data to improve tumor segmentation. Specifically, the registration process aligns both the extracted features of the ultrasound and magnetic resonance data and the original images. As a result, our loss function consists of three components: the loss of registration for the original images, the loss of similarity for the extracted features, and the loss of segmentation for the tumor. $G_M$ and $G_T$ represent the ground truth for MRI and TRUS images, where $G_M$ and $G_T \in \mathbb{R}^{3 \times W \times H \times D}$ represent the ground truth of the each voxel, corresponding to "other", "prostate" and "tumor", respectively.

\textbf{Loss for alignment}: We employ weighted dice loss to optimize the registration module which is defined as:

\begin{equation}
\mathcal{L}_{reg} = 1 - \frac{\sum_{c=1}^{3} w_c \left( 2 \sum_{i=1}^{n} (\hat{G}_{M}^{(c)} \cdot G_{T}^{(c)}) \right) + \epsilon}{\sum_{c=1}^{3} w_c \left( \sum_{i=1}^{n} (\hat{G}_{M}^{(c)})^2 + \sum_{i=1}^{n} (G_{T}^{(c)})^2 \right) + \epsilon},
\end{equation}

where $\hat{\boldsymbol{G}}_{M}$ is the transformed mask obtained by applying the transformation matrix $\boldsymbol{A}$, such that $\hat{\boldsymbol{G}}_{M} = \mathds{T}(\boldsymbol{G}_{M},\boldsymbol{A})$, $n = W \times H \times D$ represents the total number of voxels, $w_c$ denotes the weight of the class $c$, and $\epsilon$ is a small constant added to prevent division by zero. To reflect the importance of each class, we assign a higher weight to the tumor class compared to the prostate and other classes. 

\textbf{Loss for feature distribution}: KL divergence is used to measure the similarity between the feature distributions of $\hat{\boldsymbol{M}}_{merge}$ and $\boldsymbol{T}_{merge}$, ensuring that similar features are extracted from both modalities. As shown in step \circled{10} of Fig.\ref{Figure method}, the feature distributions of $\hat{\boldsymbol{M}}_{merge}$ and $\boldsymbol{T}_{merge}$ are extracted using convolutional neural networks, followed by softmax layers. These feature distributions are denoted as $\hat{\boldsymbol{M}}_{dis}$ and $\boldsymbol{T}_{dis}\in \mathbb{R}^d$, respectively. We then compute the similarity between their probability distributions. The KL divergence loss function for $\hat{\boldsymbol{M}}_{dis}$ and $\boldsymbol{T}_{dis}$, both in $d$-dimensional space, is given by:
\begin{equation}
\mathcal{L}_{KL} = \sum_{i=1}^{d} \boldsymbol{T}_{dis}(i) \log \frac{\boldsymbol{T}_{dis}(i)}{\hat{\boldsymbol{M}}_{dis}(i)},
\end{equation}
The purpose of using KL divergence here is to ensure that both $\boldsymbol{M}_{fea}$ and $\boldsymbol{T}_{fea}$ extract features with the same distribution from the original image for accurate registration. Since the convolution-based feature extractor preserves the spatial positions of the features, $\boldsymbol{A}$ is applied to adjust the spatial alignment of $\boldsymbol{M}_{fea}$'s features.
\textbf{Loss for segmentation}: 
The segmentation loss function $\mathcal{L}_{\text{seg}}$ is a combination of loss of the dice and loss of the focal. Dice loss maximizes the overlap between the predicted segmentation and the ground truth, while Focal Loss addresses class imbalance by focusing more on hard-to-classify examples.
\begin{equation}
\mathcal{L}_{seg} = \mathcal{L}_{dice} + \mathcal{L}_{focal},
\end{equation}
\begin{equation}
\mathcal{L}_{dice} = 1 - \frac{2 \sum_{c=1}^{3} \sum_{i,j,k} \left(Y^{(c)}(i,j,k) \cdot G_T^{(c)}(i,j,k) \right)}{\sum_{c=1}^{3} \sum_{i,j,k} \left( Y^{(c)}(i,j,k) + G_T^{(c)}(i,j,k) \right)},
\end{equation}
\begin{equation}
\mathcal{L}_{focal} = - \alpha \sum_{c=1}^{3} \sum_{i,j,k} \left( (1 - Y^{(c)}(i,j,k))^\gamma \log(Y^{(c)}(i,j,k)) \right),
\end{equation}

The final loss of our method is:
\begin{equation}
\mathcal{L} = \alpha\mathcal{L}_{seg} + \beta\mathcal{L}_{KL} + \lambda\mathcal{L}_{seg},
\label{eq:loss}
\end{equation}
In this equation, $\alpha$, $\beta$ and $\lambda$ is balancing factors that controls the contribution of the each loss.

\subsection{Implementation}
The training process is divided into two phases: a) the pretraining phase, which optimizes the parameters of registration, i.e., pre-training the network by using the registration loss functions, and b) the segmentation optimization phase, we take the encoder weights derived from pre-training as initial values and freeze them to optimize the decoder for segmentation. Our method is trained on a standalone workstation equipped with a Nvidia RTX A6000 GPU and an Intel Core i7-7700 CPU. We adopt the Adam optimizer with a fixed learning rate of $1e^{-4}$ and batch size sets of 1 for all learning-based approaches. To accelerate the convergence of the model during the overall training process, we conducted a 100-epoch pre-training on the registration component. This step helps to stabilize the initial learning phase and improves the model's performance when training in an end-to-end manner. 

\subsection{Evaluation Metrics}
Quantitative evaluations of various models were performed at both the lesion and patient levels. We assessed performance using metrics such as the area under the Receiver Operating Characteristic curve (ROC), sensitivity (SE), specificity (SP), negative predictive value (NPV), positive predictive value (PPV), and accuracy (ACC).

\textbf{Lesion Level}: The assessment of true positive and false negative lesions was determined based on the overlap between predicted and actual lesions. A detection was classified as a true positive if the predicted labels overlapped with at least 1\% of the actual lesion; otherwise, it was considered a false negative. To distinguish true negative and false positive lesions, the prostate was divided into six segments. A segment was classified as ground truth negative if it contained less than 1\% actual cancer voxels. Conversely, if 99\% or more of the predicted labels in a segment were normal, it was categorized as a true negative. Otherwise, it was considered a false positive.

\textbf{Patient-Level}: In the patient-level evaluation, a patient was considered a true positive if the models correctly detected at least one lesion. If no lesions were correctly identified, the patient was classified as a false negative.

\begin{table*}[h!]
\centering
\renewcommand{\arraystretch}{1.3}
\begin{tabular}{ccccccc}
\hline
\textbf{Data} & 
\textbf{Method} & Dice (\%) $\uparrow$ & AUC(\%) $\uparrow$& Sen (\%) $\uparrow$& Spe (\%) $\uparrow$& Acc (\%) $\uparrow$
\\ \hline
TRUS  & Unet &0.117	&0.617	&0.414	&0.844&0.794   
\\ \hline
\multirow{2}{*}{TRUS \& DWI}  & Unet &0.083	&0.574	&0.456	&0.710	&0.677\\ 
& Ours &\textcolor{red}{\textbf{0.227}}	&\textbf{0.761}	&\textbf{0.611}	&\textbf{0.908}	&\textbf{0.859}
\\ \hline
\multirow{2}{*}{TRUS \& ADC}  & Unet &0.118&	0.624	&0.441&	0.827	&0.782
\\ 
& Ours &\textbf{0.191}	&\textbf{0.716}	&\textbf{0.562}	&\textbf{0.885}	&\textbf{0.847}
\\ \hline
\multirow{2}{*}{TRUS \& T2}  & Unet &0.114	&0.578	&\textbf{0.590}	&0.593	&0.597
\\ 
& Ours &\textbf{0.164}	&\textbf{0.697}	&0.484	&\textcolor{red}{\textbf{0.915}}	&\textbf{0.850}
\\ \hline          
\multirow{2}{*}{TRUS \& ALL}  & Unet &0.132	&0.587	&\textcolor{red}{\textbf{0.712}}	&0.499	&0.509
\\ 
& Ours &\textbf{0.212}&\textcolor{red}{\textbf{0.779}}	&0.667	&\textbf{0.896}	&\textcolor{red}{\textbf{0.880}}
\\ \hline
\end{tabular}
\caption{Quantitative Comparison of Various Data Combinations for Training Methods. This table compares the performance of semantic segmentation models trained with ultrasound data and different types of MRI data. The ``\&'' symbol indicates that both ultrasound and MRI data were input simultaneously during model training. ``ALL'' denotes ultrasound images across three MRI sequences, including T2-weighted, ADC, and DWI.}
\label{Table1_comparision}
\end{table*}

\section{Results}
\begin{figure*}
    \centering
    \includegraphics[width=0.85\linewidth]{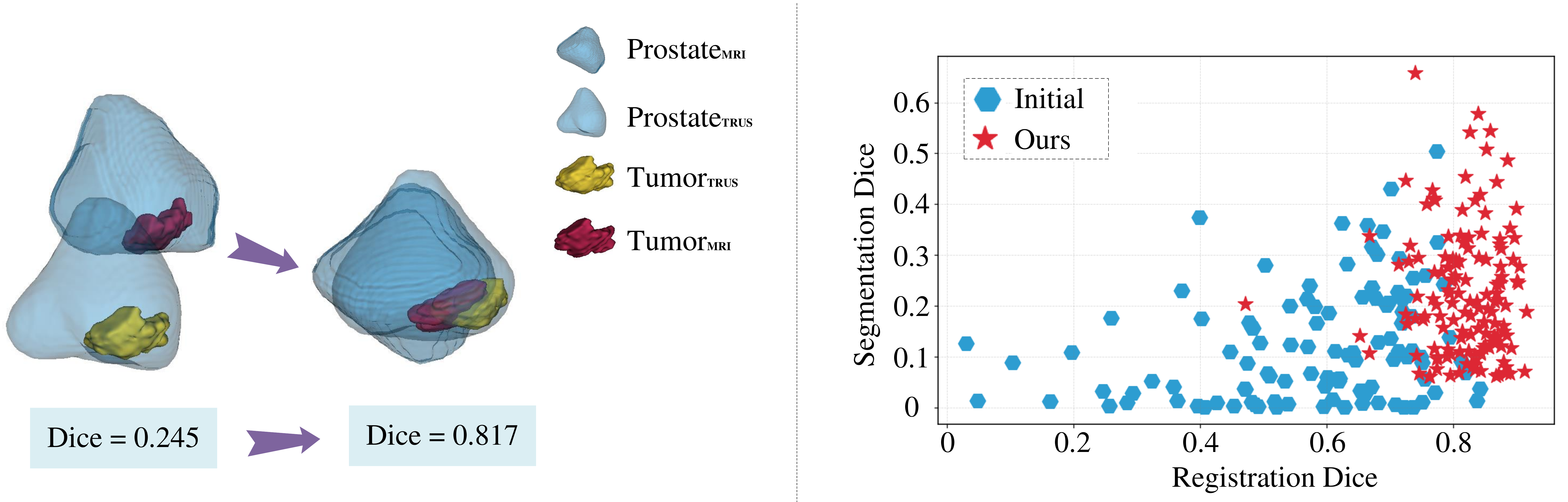}
    \caption{Effect of TRUS-MRI Alignment on Prostate and Tumor Segmentation Performance.The left panel illustrates the spatial positions of the prostate and tumor in the initial TRUS and MRI inputs, along with an intermediate output from our registration method. The right panel depicts the relationship between the alignment quality of the input data (TRUS and MRI) and the final model performance. The red points represent intermediate outputs from our registration approach. As shown, our method significantly aligns the TRUS and MRI data compared to the initial input, resulting in an improvement in model performance.}
    \label{Figure_dice_between_mri_trus}
\end{figure*}

\subsection{Quantitative Evaluation and  Analysis}
Current multimodal techniques typically train models by concatenating all available data at various stages \cite{karani2022review,stahlschmidt2022multimodal,gadzicki2020early,boulahia2021early}. 
We evaluated the effectiveness of training models for prostate cancer segmentation by combining ultrasound and different MRI modalities. Table \ref{Table1_comparision} presents the experimental results obtained by training models directly using ultrasound images and different types of MRI data, as well as the results from our proposed method. From the table, it can be observed that:

1) Training models directly with combined ultrasound and magnetic resonance data does not yield better performance than using ultrasound data alone, indicating that incorporating more data does not necessarily lead to improved results. For example, Table\ref{Table1_comparision} illustrates the performance of the Unet model\cite{ronneberger2015u} trained with identical parameter settings using TRUS alone and TRUS combined with MRI (denoted as TRUS\&MRI) simultaneously for the prostate tumor segmentation task. Although the ``TRUS\&MRI'' utilizes more data for model training, there are no significant improvements for model performance. This is because segmentation tasks are sensitive to spatial information, and the prostate's position in MRI and ultrasound data differs significantly. This discrepancy causes the prostate-related information in the MRI data to interfere with the model, introducing noise-like artifacts. As illustrated in Fig.\ref{Figure_dice_between_mri_trus}a, the prostate and cancer regions in the initial TRUS and MRI data are misaligned. When MRI and TRUS data are combined and fed into the model for training, the regions marked as cancer on MRI labels may not correspond to cancer in the TRUS data, leading to noise and misguided direction. Table\ref{Table_dice_prostate_tumor} presents statistics on prostate and tumor regions in the initial TRUS and MRI data, with prostate metrics of 0.597 and 0.071, respectively. These discrepancies highlight the spatial misalignment of prostate-related information, which hinders performance improvements when directly using additional data. 

2) Our method significantly improves the model's ability to identify prostate tumor using multimodal data. Compared to the Unet model, our approach achieves substantial improvements in all metrics for various combinations of modality. For example, while the Unet model trained on TRUS and multisequence MRI data achieved a dice score of 0.132, our method improved this score to 0.212 - an overall average improvement of 92\% for the dice metric. Similar improvements were observed in other evaluation metrics. This is because our method does not directly train the segmentation model on multimodal data but instead aligns the different modalities within the model before segmentation. Figure\ref{Figure_dice_between_mri_trus}b illustrates the alignment of the prostate regions in the initial ultrasound and magnetic resonance data and the intermediate results produced by our registration module. The improved alignment achieved by our method enhances the ability of the segmentation module to accurately segment tumors.

The above results demonstrate the importance of modality alignment in using multimodal data for prostate tumor detection, providing a robust solution to mitigate the challenges posed by spatial discrepancies in heterogeneous data sets.
\begin{table*}[h!]
\centering
\renewcommand{\arraystretch}{1.3}
\begin{tabular}{ccccccccc}
\hline
\multirow{2}{*}{\textbf{Method}} & \multicolumn{2}{c}{\textbf{Registration}} & \multicolumn{5}{c}{\textbf{Segmentation}} 
\\ 
\cline{2-8} 
& Dice (\%) $\uparrow$& HD (mm) $\downarrow$ & Dice (\%) $\uparrow$ & AUC(\%) $\uparrow$ & Sen (\%) $\uparrow$ & Spe (\%) $\uparrow$ & Acc (\%) $\uparrow$                             
\\ \hline
Initial & 0.597 & 16.721 &0.132	&0.587	&\textbf{0.712} &0.499	&0.509
\\ 
ANTs\cite{avants2009advanced}  & 0.252 & 19.990 & 0.115 	& 0.597 	& 0.445 & 0.581 & 0.701  
\\
ConvNet\cite{de2019deep}
& 0.771 & 9.002 & 0.157	&0.731	& 0.340	& 0.891	& 0.897
\\
VTN-Affine\cite{zhao2019unsupervised} & 0.752 &  \textbf{8.758} & 0.169	& 0.711	& 0.532	& 0.896	& 0.881
\\
C2FViT\cite{mok2022affine} & \textbf{0.814} & 8.787 & 0.201	& 0.704	&	0.529 & 0.884	& \textbf{0.904}
\\ 
Ours (Independent) & 0.779 & 8.835 &  0.178	&0.700	&0.520	&0.890	&0.870
\\
Ours (End to End) & 0.807 & 8.791 & \textbf{0.212} & \textbf{0.779}	& 0.667	& \textbf{0.896}	& 0.880
\\ \hline
\end{tabular}
\caption{Segmentation and Registration Results Comparison. ↑: higher is better, and ↓: lower is better. Initial: initial results in native space without registration.}
\label{Table2_registration_segmentaion}
\end{table*}
\subsection{Decoupling the Segmentation and Registration Steps}
The analysis of the results in Table\ref{Table2_registration_segmentaion} and Fig.\ref{Figure_dice_between_mri_trus} indicates that the alignment of similar information significantly affects the performance of models trained on multimodal data. In response to this observation, we analyzed the performance of a segmentation model trained using aligned multimodal data. The experimental pipeline was conducted in two stages. First, the ultrasound and MRI data were aligned based on the prostate gland in the images. The aligned ultrasound and magnetic resonance data were then used to train the semantic segmentation model. This experiment had two main objectives: 1) to analyze the impact of data alignment (via registration methods) on segmentation performance, and 2) to evaluate the advantages of our method compared to the stage-wise approach for registration and segmentation.

For the registration methods, we employed the widely used ANTs\cite{avants2009advanced} method along with three deep learning-based registration approaches: ConvNet\cite{de2019deep}, VTN-Affine\cite{zhao2019unsupervised} and C2FViT\cite{mok2022affine}. ANTs\cite{avants2009advanced} was selected as a standalone registration module because it is a representative optimization-based registration method, while the other three methods are state-of-the-art deep learning-based approaches. These methods were chosen to evaluate how registration techniques based on different principles influence the segmentation results. Ours (Independent) refers to our method in which the registration and segmentation modules are trained separately.
The overall experimental results are summarized in Table\ref{Table2_registration_segmentaion}. The following observations can be drawn from the table:

1) The alignment of information in multimodal data significantly enhances the performance of segmentation models. For example, training segmentation models using multimodal data aligned by ConvNet\cite{de2019deep}, VTN-Affine\cite{zhao2019unsupervised} and C2FViT\cite{mok2022affine}, and Ours (independent) led to substantial improvements in the model’s ability to identify tumors in ultrasound data. Specifically, compared to directly training the model on the original data, which achieved a tumor recognition Dice score of 0.132, the aligned data obtained using ConvNet\cite{de2019deep}, VTN-Affine\cite{zhao2019unsupervised} and C2FViT\cite{mok2022affine} improved the Dice score by 18.94\%, 28.03\%, and 52.27\%, respectively. These results demonstrate that aligning MRI data with ultrasound data based on the prostate gland significantly enhances the model's ability to identify prostate tumors from ultrasound data alone. This finding underscores the critical role of multimodal data alignment in improving segmentation performance.

\begin{table}[h!]
\centering
\renewcommand{\arraystretch}{1.3}
\begin{tabular}{p{2cm}p{1.5cm}p{1.5cm}}
\hline
\multirow{2}{*}{\textbf{Method}} & \multicolumn{2}{c}{\textbf{Registration}} 
\\ 
\cline{2-3} 
& Dice$_{1}$(\%)& Dice$_{2}$(\%)           
\\ \hline
Initial & 0.597 & 0.071 
\\ 
ANTs\cite{avants2009advanced}  & 0.252 & 0.003 
\\
ConvNet\cite{de2019deep}  & 0.771 & 0.359 	
\\
VTN\cite{zhao2019unsupervised} & 0.752 & 0.321 
\\
C2FViT\cite{mok2022affine} & \textbf{0.814} & 0.423
\\ 
Ours$_{Ind}$ & 0.779 & 0.390 
\\
Ours$_{E2E}$ & 0.807 & \textbf{0.464}
\\ \hline
\end{tabular}
\caption{Visual registration results using different approaches. Each row represents a different registration method applied to the same prostate, and the columns correspond to five evenly spaced slices of the prostate. The slices were selected by taking the first and last slices of the prostate region from the ultrasound images and dividing the interval into five equal parts.}
\label{Table_dice_prostate_tumor}
\end{table}

2) Simultaneous training of segmentation and registration modules can improve the performance of both tasks. As shown in Table\ref{Table2_registration_segmentaion}, when the registration and segmentation modules of our method are trained separately, the registration module achieves a performance of 0.779, while the corresponding segmentation performance is 0.169. In contrast, simultaneous training improves the registration performance to 0.817 and significantly increases the segmentation performance to 0.212. The models used in stage-wise training and simultaneous training are identical, with the only difference being that simultaneous training employs an end-to-end approach. In this approach, registration and segmentation are integrated through steps 6 and 10, as described in the section\ref{method:regsegnet}, and the loss function outlined in Equation\ref{eq:loss}. Specifically, the registration module aligns MRI data with ultrasound data for the segmentation module, while the segmentation module's predictions of the prostate and tumor are fed back to refine the registration process. This joint training framework allows the registration and segmentation modules to mutually enhance each other's performance, highlighting the effectiveness of integrating these tasks.

3) Higher prostate alignment may not always result in improved performance in predicting prostate tumors. For example, although C2FViT\cite{mok2022affine} achieves the highest registration performance of 0.814, slightly exceeding our method's performance of 0.807, our method achieves a tumor segmentation performance of 0.212, which surpasses C2FViT's corresponding performance of 0.201. This discrepancy underscores that prostate registration performance is not linearly correlated with tumor segmentation performance. This arises from the focus of registration methods adopted in the experiment, which primarily align prostate gland data between ultrasound and MRI without ensuring improved alignment of the associated tumor information. To investigate this, we analyzed the prostate gland and prostate tumor registration results, as presented in Table\ref{Table_dice_prostate_tumor}. Although our method slightly underperforms C2FViT in prostate gland registration, it achieves a superior tumor alignment between ultrasound and MRI. This enhanced tumor alignment, being more closely associated with the segmentation task, significantly boosts the performance of the tumor segmentation model trained on multimodal data. The above results indicate that aligning the most task-relevant data is crucial when training multimodal models to enhance segmentation performance.

\begin{figure}
    \centering
    \includegraphics[width=1\linewidth]
    {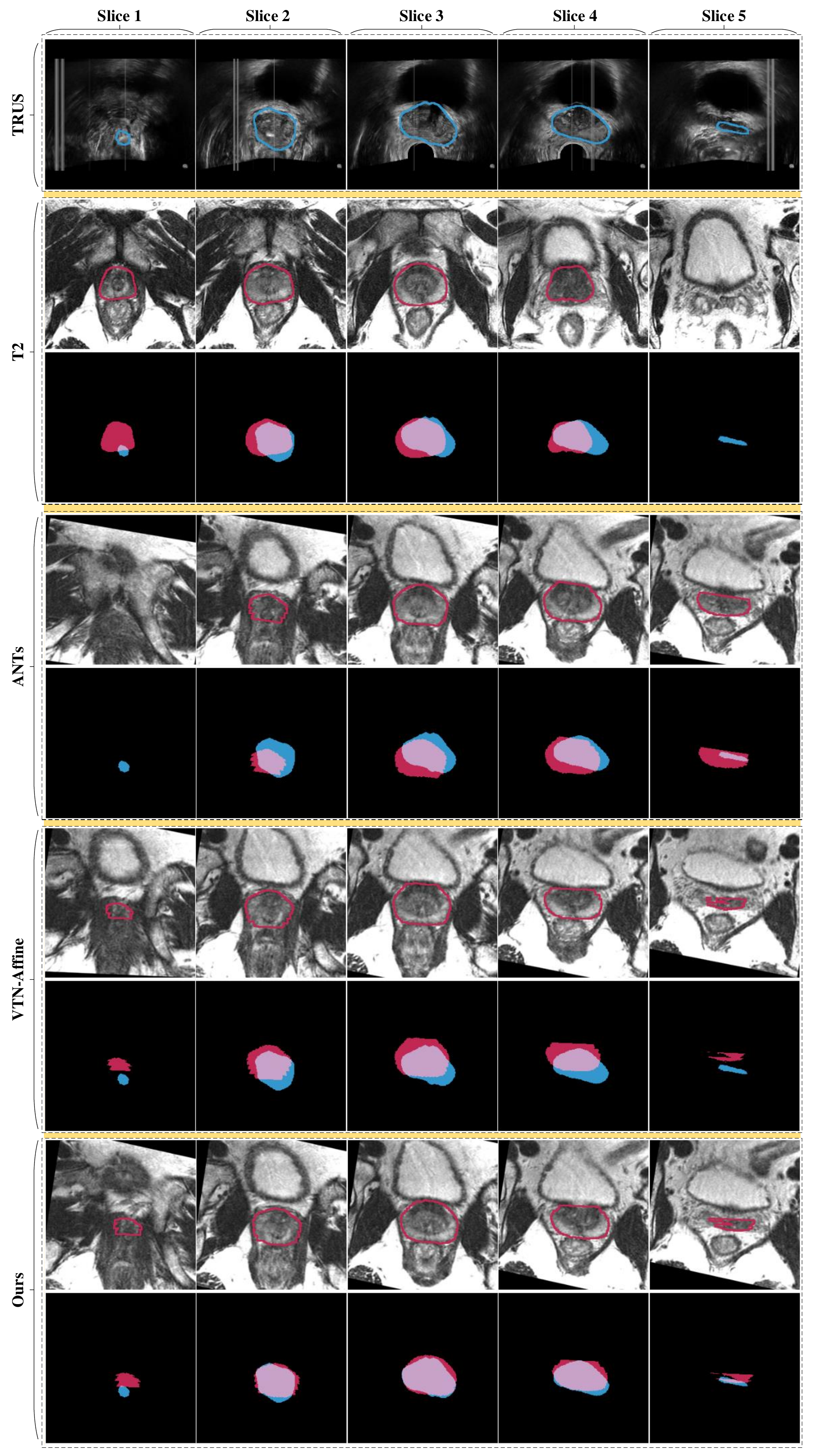}
    \caption{Visualization of registration and segmentation results across different training epochs. The top row shows the registration results, with the Dice score for registration performance displayed below each 3D model. The bottom row illustrates the segmentation results, highlighting the segmented tumor regions in color and the corresponding Dice score for segmentation performance. The results are shown at selected epochs to demonstrate the progressive improvement of both registration and segmentation during training. The comparison highlights the relationship between training progression and model performance for both tasks.
    }   
    \label{Figure_result_of_registration}
\end{figure}
In addition, C2FViT\cite{mok2022affine} achieves the highest performance among registration methods, primarily because it adopts a three-layer structure incorporating varying granularity levels in the registration process. In contrast, our method uses a single-layer registration module to enhance training efficiency. Additionally, ANTs\cite{avants2009advanced} shows even worse prostate alignment performance compared to the original data. This is likely due to the significant morphological differences between prostate ultrasound and MRI images. Since ANTs\cite{avants2009advanced} relies on similarity-based matching, its effectiveness is significantly reduced when addressing such significant morphological discrepancies. These findings highlight the inherent challenges of using multimodal prostate data for prostate tumor segmentation.

Based on the above analysis, our method enhances the model's ability to identify prostate tumors in ultrasound by aligning prostate-related information between ultrasound and MRI. It surpasses the stage-wise training approach, which first registers the data and then uses the aligned data for segmentation training. The strength of our method lies in its simultaneous registration and segmentation training, allowing the two processes to mutually enhance each other's results.

%
%

\subsection{Qualitative Evaluation of Registration and Segmentation Results}
Our proposed method improves the model's ability to identify prostate tumors from TRUS by aligning prostate-related information across multiple modality data. In our framework, we adopt an affine registration approach to align information. Although the model's primary objective is to identify prostate tumors in TRUS and not to explicitly output multimodal registration results, the registration process is a critical component of the model. Here, we show and analyze the prostate registration results (intermediate outputs) and the prostate tumor segmentation results.

Figure\ref{Figure_result_of_registration} shows the registration results for TRUS and T2-weighted MRI data. Due to space constraints, only the outcomes of the optimization-based method (ANTs\cite{avants2009advanced}) and the deep learning method (VTNAffine\cite{zhao2019unsupervised}) are displayed, with additional results provided in Supplementary Material 1. Since prostate imaging involves 3D data, the registration is performed in three dimensions. To better demonstrate this, five representative slices of TRUS and MRI images are presented, evenly spaced along the prostate's spatial extent from the first to the last slice containing the prostate. From Figure\ref{Figure_result_of_registration} we can know that: 1) Significant initial spatial divergence: The unregistered TRUS and MRI data show substantial spatial misalignment of the prostate gland. For example, the prostate regions in the TRUS and T2-weighted images are significantly misaligned across the sagittal, axial, and coronal planes. 2) Challenges in multimodal prostate registration: Unlike multimodal imaging of organs such as the brain or spine, where anatomical features remain relatively consistent across modalities, the appearance of the prostate in TRUS and MRI differs significantly. This substantial disparity in modality-specific characteristics, as discussed in the previous section, presents a significant challenge for traditional registration methods like ANTs\cite{avants2009advanced} and underscores the difficulty of achieving accurate multimodal registration for the prostate. 3) Our method demonstrates improved registration performance compared to previous approaches, particularly in aligning the peripheral edges of the prostate. For example, while VTN enhances alignment in the central slices, it fails to accurately align the prostate edges, as observed in slices 1 and 5. Accurate alignment of the peripheral prostate is essential for tumor detection, as tumors are more frequently located in the peripheral zone\cite{haffner2009peripheral,sakai2005comparison}. In contrast, our method successfully aligns these critical edge regions, enabling more reliable tumor segmentation.

\begin{figure}
    \centering
    \includegraphics[width=1\linewidth]{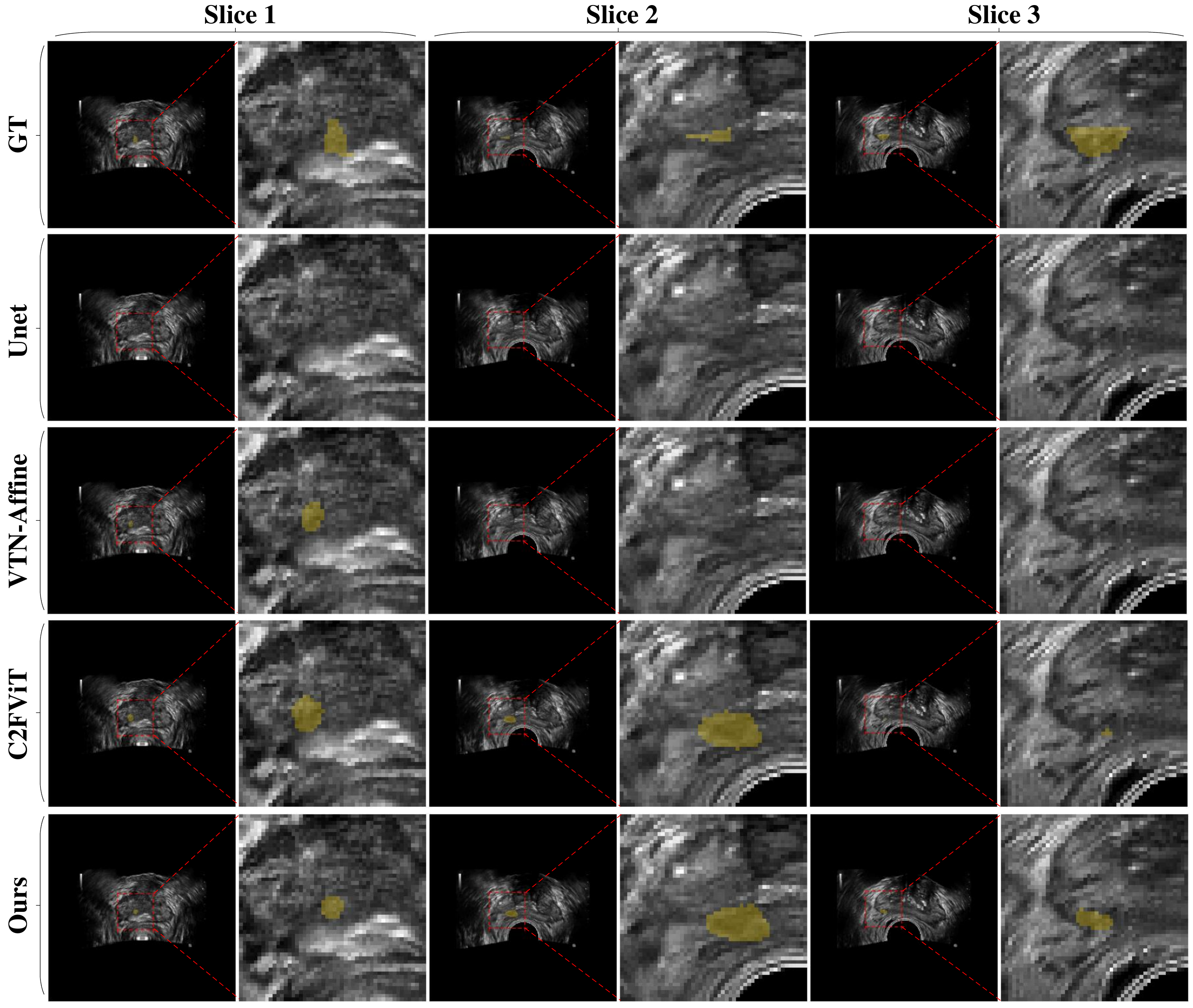}
    \caption{Segmentation results across different methods. Each row represents the segmentation results generated by a specific method, and the columns correspond to three selected slices of the prostate. The yellow regions indicate the segmented prostate tumors, while the red boxes highlight the regions of interest for better visualization. }
    \label{Results of segmentation}
\end{figure}

Figure\ref{Results of segmentation} presents the prostate tumor segmentation results on TRUS images from different models. To provide a more comprehensive visualization of tumor detection in 3D TRUS images, we include three representative slices: the first and last slices containing the prostate tumor and a middle slice between them. Figure\ref{Results of segmentation} shows that: 1) Limited segmentation performance without registration: The model trained on unregistered TRUS and MRI data achieves the lowest segmentation performance, failing to detect tumors in TRUS images. This limitation arises from the inability of unaligned MRI data to effectively support tumor identification in TRUS, an inherently challenging task, as evidenced by commonly reported Dice scores of approximately 0.11 \cite{vesal2022domain,bhattacharya2022review}. 2) Enhancing tumor segmentation through registration: The results demonstrate that aligning TRUS and MRI data through registration significantly improves the model's ability to detect prostate tumors in TRUS images. For instance, the Unet trained on VTN-registered data detected the tumor in slice 1, while the Unet trained on C2FViT-registered data identified tumors in slices 1 and 2. As shown in Table\ref{Table2_registration_segmentaion}, C2FViT\cite{mok2022affine} achieves higher registration accuracy than VTN\cite{avants2009advanced}, contributing to its slightly better segmentation performance. Moreover, our proposed method accurately detects tumors in all slices, outperforming other approaches in accuracy and consistency. The ability to detect tumors across all slices suggests that our method is more robust in handling challenging cases and provides more complete segmentation outcomes than alternative methods.

These results emphasize the significance of multimodal alignment in improving segmentation, especially in challenging scenarios such as prostate tumor detection in TRUS.

\begin{figure*}
    \centering
    \includegraphics[width=0.75\linewidth]{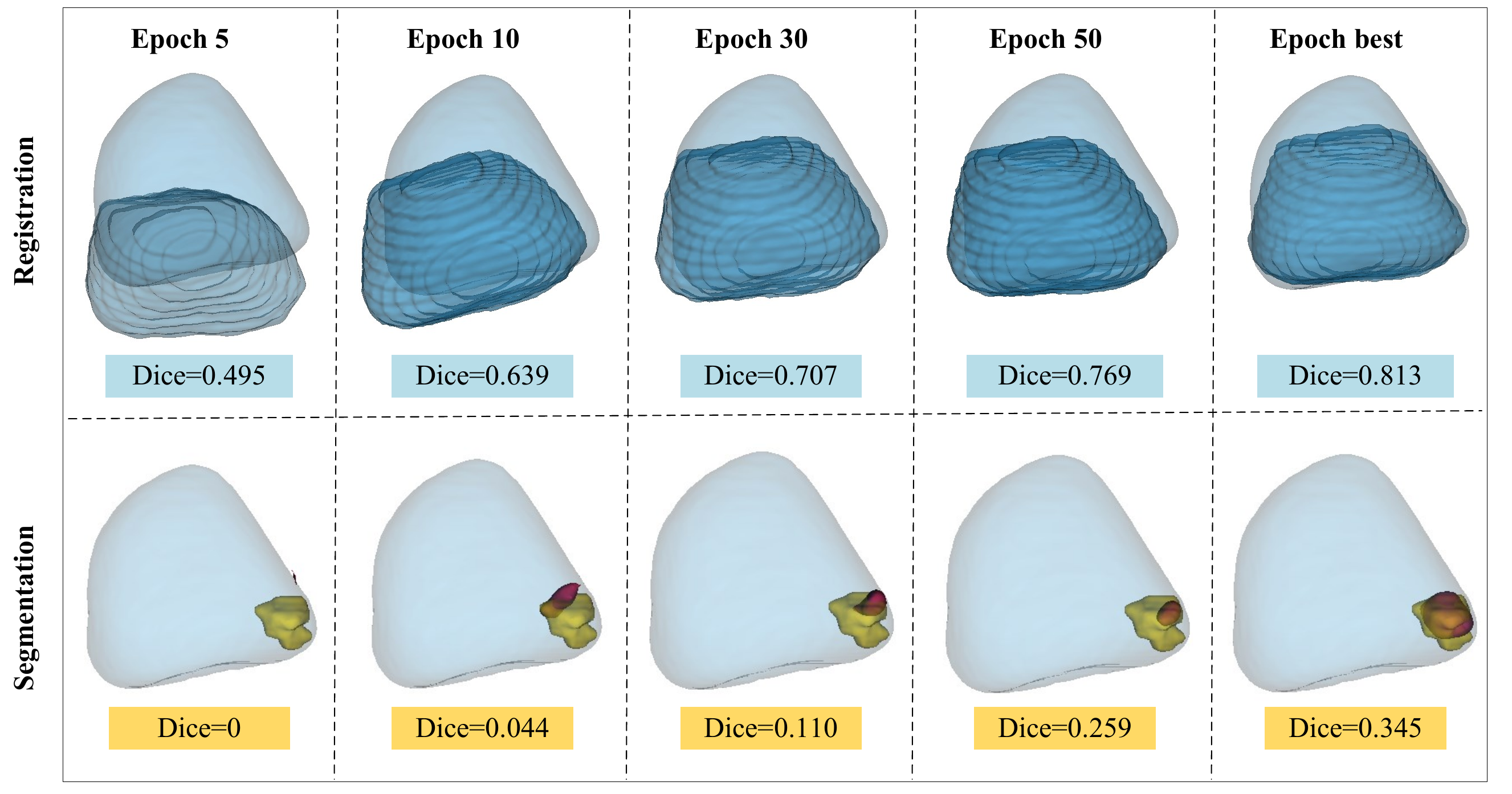}
    \caption{Visualization of registration and segmentation results across different training epochs. The top row shows the registration results, with the Dice score for registration performance displayed below each 3D model. The bottom row illustrates the segmentation results, highlighting the segmented tumor regions in color and the corresponding Dice score for segmentation performance. The results are shown at selected epochs to demonstrate the progressive improvement of both registration and segmentation during training. The comparison highlights the relationship between training progression and model performance for both tasks.}
    \label{Figure_results_reg_seg_epochs}
\end{figure*}
\subsection{Epochs between Registration and Segmentation}
We investigated the interaction between our method's registration and segmentation modules during the training process. Figure\ref{Figure_results_reg_seg_epochs} illustrates the model's registration and segmentation performance across different training epochs. The results shown in the figure represent the model's outputs on the same test dataset, with each epoch \# marking the specific checkpoint during training when the corresponding model was saved.

As shown in Fig.\ref{Figure_results_reg_seg_epochs}, the performance of both the registration and segmentation modules improves simultaneously as training progresses. For instance, during the early stages of training (epoch 5), there is significant spatial misalignment between the prostate in the TRUS and MRI data, and the model fails to detect prostate tumors in the TRUS image. As training continues, registration performance gradually improves, with the alignment metric increasing from 0.495 at epoch 5 to 0.769 at epoch 50. At the same time, the segmentation capability also improves, with the model beginning to detect tumors at epoch 10 and segmentation performance progressively increasing from 0.044 at epoch 10 to 0.259 at epoch 50. These results demonstrate that, in our method, the registration and segmentation modules are jointly optimized during training. The improvement in tumor segmentation performance is directly correlated with the enhanced alignment capability of the model, emphasizing the importance of accurate registration for achieving effective tumor segmentation in multimodal data.

Specifically, the registration module's output serves as input to the segmentation model, which, in turn, guides and constrains the registration module's learning process. This interdependent relationship enhances segmentation accuracy by improving spatial consistency through registration, while segmentation provides additional structured information to guide the registration model toward more accurate anatomical alignment. In contrast, most joint registration-segmentation methods employ two independent encoder-decoder structures for registration and segmentation tasks, training these components separately. This differs from our approach, where we integrate the registration method into the segmentation pipeline and explore their combined effects on segmentation performance. By incorporating registration as an integral part of the segmentation process, we achieve a more unified and effective training workflow that leverages the strengths of both modules.

\section*{Discussion}
\textit{Why adopted affine registration for data alignment?} Common registration methods include rigid registration, affine registration, and deformable registration. We selected affine registration primarily because it offers better performance compared to rigid registration while requiring significantly less computational resources than deformable registration. Specifically, affine registration computes a 4×4 transformation matrix to align the spatial positions of MRI data with TRUS, which can then be used for training the segmentation model. In contrast, deformable registration generates a deformation field that is three times the size of the MRI data volume, significantly increasing computational complexity when aligning MRI with TRUS for subsequent model training. The efficiency of obtaining a compact 4×4 transformation matrix makes affine registration a practical choice in this work. Another key reason for this choice is the primary goal of this study: to explore how multimodal data alignment can improve prostate cancer detection from ultrasound images. Our experimental results demonstrate that multimodal data alignment is crucial for training effective segmentation models. While affine registration was used in this study, exploring alternative registration methods, such as deformable registration, remains an important direction for future research.

\textit{What other information beyond the prostate can be utilized for multimodal data alignment?}
Our experimental results also reveal that aligning prostate data from TRUS and MRI improves the model’s ability to detect prostate tumors. Furthermore, the enhanced alignment provided by our method leads to better segmentation performance, suggesting that aligning multimodal data relevant to prostate cancer significantly benefits tumor detection. In this work, all registration methods were trained using labeled TRUS and MRI data. This raises an intriguing question: could a model autonomously identify and align the most relevant features for prostate cancer detection in an unsupervised manner? Enabling models to automatically identify and align features most critical for prostate cancer detection from multimodal data could further improve segmentation accuracy and provide deeper insights into the key features of prostate cancer. Developing unsupervised learning approaches for feature-based alignment is a promising direction for future research.

\section*{Acknowledgments}
This work was supported by  Stanford University (Departments: Radiology, Urology) and by National Cancer Institute, National Institutes of Health (R37CA260346). The content is solely the responsibility of the authors and does not necessarily represent the official views of the National Institutes of Health.

\bibliographystyle{elsarticle-num-names}

\end{document}